\RequirePackage[2020-02-02]{latexrelease}
\documentclass[twocolumn,english,aps,prl,10pt,superscriptaddress]{revtex4}

\usepackage[T1]{fontenc}
\usepackage{amsmath,amssymb,graphicx,bm,color}

\begin{document}

\title{Emergence of Nonlocal Hydrodynamics in Many-Body Radiative Heat Transport}

\author{M. G\'omez Viloria}
\affiliation{Laboratoire Charles Fabry, UMR 8501, Institut d'Optique, CNRS,
Universit\'e Paris-Saclay, 2 Avenue Augustin Fresnel,
91127 Palaiseau Cedex, France}

\author{F. Herz}
\affiliation{Laboratoire Charles Fabry, UMR 8501, Institut d'Optique, CNRS,
Universit\'e Paris-Saclay, 2 Avenue Augustin Fresnel,
91127 Palaiseau Cedex, France}

\author{R. Messina}
\affiliation{Laboratoire Charles Fabry, UMR 8501, Institut d'Optique, CNRS,
Universit\'e Paris-Saclay, 2 Avenue Augustin Fresnel,
91127 Palaiseau Cedex, France}

\author{P. Ben-Abdallah}
\email{pba@institutoptique.fr}
\affiliation{Laboratoire Charles Fabry, UMR 8501, Institut d'Optique, CNRS,
Universit\'e Paris-Saclay, 2 Avenue Augustin Fresnel,
91127 Palaiseau Cedex, France}

\date{\today}

\begin{abstract}
We develop a nonlocal hydrodynamic theory of radiative heat transport
in many-body systems by coarse graining their microscopic
electrodynamic interactions. Long-range radiative exchange naturally
generates a spatially nonlocal constitutive relation in which the heat
current at a given position depends on temperature gradients over an
extended region, while energy conservation retains its local
conservative form. Fourier transport emerges as a controlled
long-wavelength limit when the interaction kernel has a finite second
spatial moment. When this moment fails to converge, spatial
nonlocality survives coarse graining and gives rise to non-Fourier
transport. The resulting framework establishes a systematic bridge
between discrete many-body fluctuational electrodynamics and a
continuous description of radiative heat transport.
\end{abstract}

\maketitle

Fluctuational electrodynamics provides the microscopic framework for describing radiative heat exchange generated by thermally fluctuating electromagnetic currents~\cite{Rytov,PoldervH,JoulainSurfSciRep05,VolokitinRMP07}. In many-body systems composed of mutually interacting thermal emitters, radiative exchange can be formulated within a Landauer-like framework in terms of electromagnetic transmission coefficients connecting spatially separated constituents~\cite{PBAEtAl2011,Riccardo,RMP}. This microscopic description has revealed collective and long-range transport regimes characterized by anomalous, or non-Fourier, thermal spreading~\cite{PRL_superdiff,Kruger,Zhao}. How the microscopic radiative transfer network determines its corresponding continuum transport kernel, and under which conditions that kernel reduces to a local conductivity, remain open questions. In conventional heat conduction, coarse graining leads to Fourier's law, whereby the heat current at a given position is determined by the local temperature gradient. Radiative transport is qualitatively different: energy can be exchanged directly between spatially separated constituents and is therefore intrinsically nonlocal. Consequently, coarse graining a many-body radiative network does not necessarily lead to a local constitutive relation. A central question is thus under which conditions a scale-independent thermal conductivity emerges from the underlying radiative interactions, and which continuum description is required when long-range transfer channels prevent such a local limit.

In this Letter, we derive a hydrodynamic theory of many-body radiative heat transport directly from the microscopic radiative conductance network. Here, hydrodynamic refers to a coarse-grained continuum
description of the conserved energy field, in which the microscopic radiative interactions are encoded in a constitutive relation for the heat current. Following the logic of spatial volume averaging and
microscopic pair-interaction current constructions~\cite{Whitaker,Irving}, we construct the coarse-grained heat current and show that its constitutive relation is intrinsically spatially nonlocal. The resulting two-point transport kernel retains explicit information on the strength, range, and geometry of the underlying radiative channels. Fourier's law is not assumed, but emerges as a controlled long-wavelength limit when the second spatial moment of the microscopic transfer network converges. When this moment fails to converge, long-range radiative interactions survive corse graining, giving rise to genuine spatial nonlocality at the hydrodynamic level.

Consider a system of $N$ reciprocal thermal emitters located at positions $\bm r_i$, each characterized by a uniform temperature $T_i$ and heat capacity $C_i$. Within fluctuational electrodynamics, the spectral radiative power exchanged between emitters $i$ and $j$ is~\cite{PBAEtAl2011}
\begin{equation}
\phi_{ij}(\omega)
=
\frac{\mathcal T_{ij}(\omega)}{2\pi}
\left[
\Theta(\omega,T_j)-\Theta(\omega,T_i)
\right],
\label{eq:flux}
\end{equation}
where $\mathcal T_{ij}(\omega)$ is the electromagnetic transmission coefficient and
\begin{equation}
\Theta(\omega,T)
=
\frac{\hbar\omega}
{\exp[\hbar\omega/(k_{\rm B}T)]-1}
\end{equation}
is the mean energy of a Planck oscillator. Linearizing the thermal balance around a uniform temperature $T_0$, with $\delta T_i=T_i-T_0$, the pairwise radiative thermal conductance reads
\begin{equation}
G_{ij}
=\int_0^\infty
\frac{d\omega}{2\pi}
\mathcal T_{ij}(\omega)
\left.
\frac{\partial\Theta}{\partial T}
\right|_{T_0}.
\label{eq:G}
\end{equation}
Introducing the transfer rates $\Gamma_{ij}=G_{ij}/C_i$, where $C_i$ is the total heat capacity of particle \(i\), the linearized temperature dynamics takes the master (energy balance) equation form
\begin{equation}
\dot{\delta T_i}
=\sum_{j\neq i}
\Gamma_{ij}
(\delta T_j-\delta T_i).
\label{eq:master}
\end{equation}
The microscopic thermal dynamics can thus be viewed as transport on a weighted spatial network, in which the rates $\Gamma_{ij}$ define direct radiative energy-transfer channels between spatially separated emitters.

\begin{figure}[t]
\centering
\includegraphics[width=0.45\textwidth]{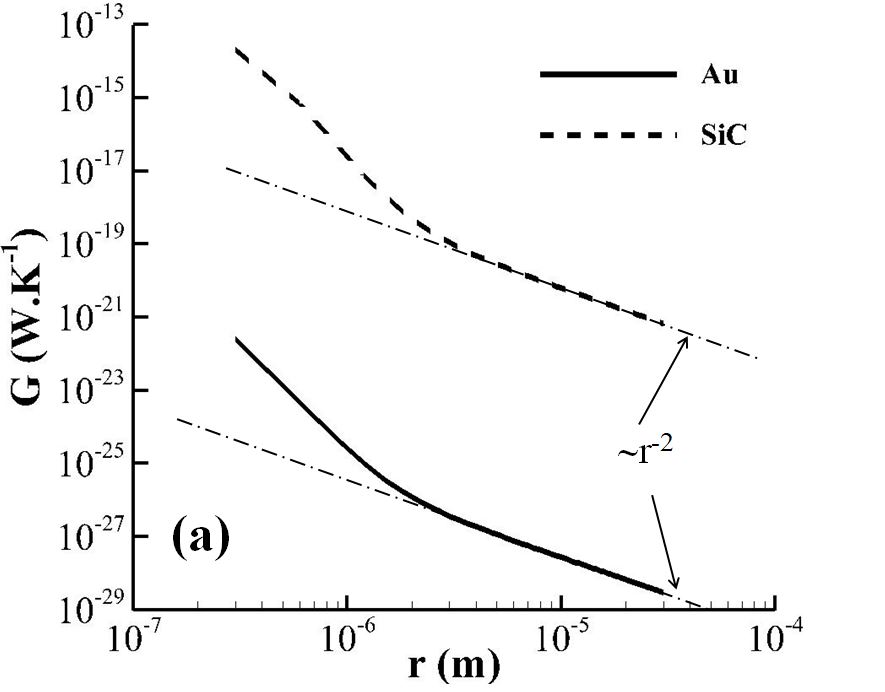}
\includegraphics[width=0.45\textwidth]{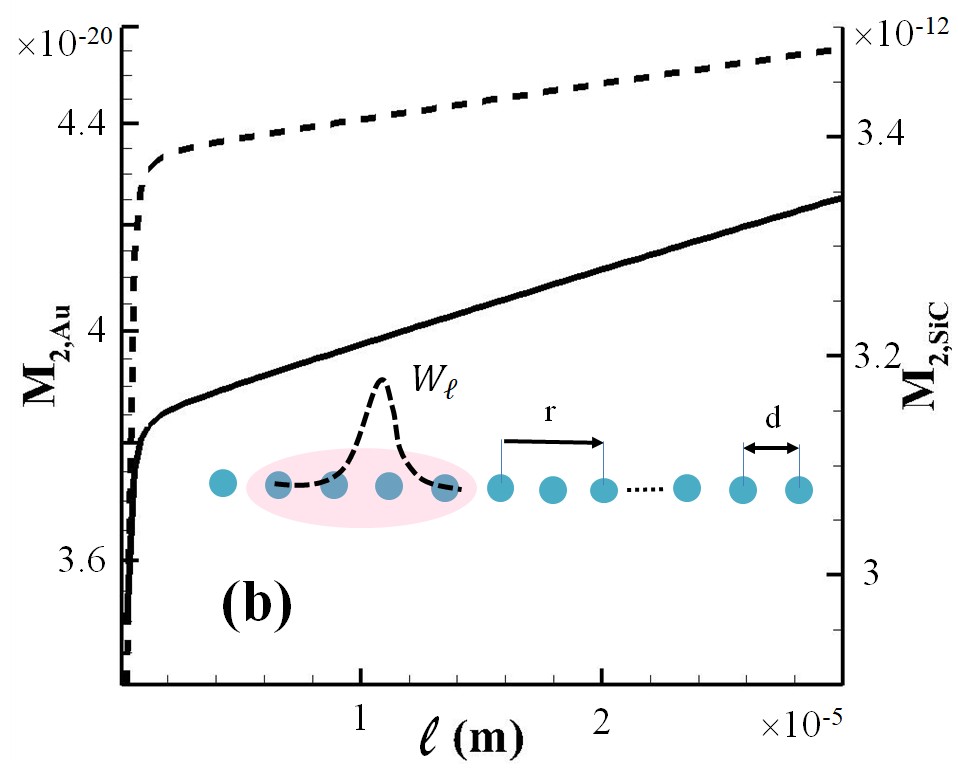}
\caption{Nonlocal transport regime in one-dimensional Au and SiC nanoparticle chains (201 nanoparticles $R=50\,$nm radius, separated by $d=4R$). (a) Pairwise radiative thermal conductance $G(r)$ as a function of the separation distance $r$ along the chain from the particle central particle. The asymptotic scaling in $r^{-2}$ illustrates the superdiffusion along the chain. (b) REV-dependent second spatial moment $\mathbb{M}_2(\ell)$ as a function of the coarse-graining length $\ell$ for the same chains. The linear scaling on $\ell$ reveals the persistence of long-range radiative channels over the considered length scales.}
\label{fig:M2}
\end{figure}

To construct a continuum description of this discrete network, we coarse grain the microscopic temperature field over a representative elementary volume (REV)~\cite{Whitaker} of characteristic size $\ell$, defining
\begin{equation}
T_\ell(\bm r,t)
=
\frac{
\sum_i W_\ell(\bm r-\bm r_i)T_i(t)}
{\sum_i W_\ell(\bm r-\bm r_i)}.
\label{eq:REV}
\end{equation}
Here $W_\ell$ is a smoothing kernel (see inset of Fig.~\ref{fig:M2}) of width $\ell$ with
\begin{equation}
\int d\bm r\,W_\ell(\bm r)=1,
\label{eq:Wnorm}
\end{equation}
which defines the spatial resolution of the coarse-grained fields. Its precise shape is not essential provided that it is sufficiently smooth and localized. For the numerical calculations we use a Gaussian kernel, which in $d$ dimensions reads
\begin{equation}
W_\ell(\bm r)
=
\frac{1}{(2\pi\ell^2)^{d/2}}
\exp\!\left(-\frac{|\bm r|^2}{2\ell^2}\right).
\end{equation}
A meaningful separation between microscopic and macroscopic scales requires
\begin{equation}
d_0\ll\ell\ll L_T,
\qquad
L_T\sim\frac{T}{|\nabla T|},
\label{eq:scales}
\end{equation}
where $d_0$ denotes the characteristic inter-emitter spacing and $L_T$ the scale over which the coarse-grained temperature varies.

Following the spirit of Irving-Kirkwood constructions~\cite{Irving}, we now construct the continuum current associated with the discrete radiative network. Consider a single channel between two points $(i,j)$ and define $\bm r_{ij}=\bm r_j-\bm r_i$. The linearized power exchanged between these point reads
\begin{equation}
P_{ij}=G_{ij}(T_j-T_i).
\label{eq:bondpower}
\end{equation}
Introducing a smooth microscopic interpolating field $T(\bm r,t)$ satisfying
\begin{equation}
T(\bm r_i,t)=T_i(t),
\end{equation}
the temperature difference between $i$ and $j$ can then be written exactly as
\begin{equation}
T_j-T_i
=
\int_0^1ds\,
\bm r_{ij}\cdot
\nabla T(\bm r_i+s\bm r_{ij},t).
\label{eq:bondgradient}
\end{equation}
This relation expresses the thermodynamic driving force associated with the radiative channel in terms of the temperature gradient sampled continuously along the segment joining the two emitters.  We next distribute the current carried by each microscopic transfer channel over its connecting segment and subsequently smooth the temperature gradient over the same channel. More precisely, a bond $(i,j)$ is represented by the segment $\bm r_i+s\bm r_{ij}$, with $s\in[0,1]$, and its contribution to the continuum current at position $\bm r$ is weighted by $W_\ell(\bm r-\bm r_i-s\bm r_{ij})$. The thermodynamic force driving this transfer channel is obtained in the same way by sampling the coarse-grained temperature gradient along the bond. The corresponding coarse-grained bond current density is
\begin{equation}
\begin{aligned}
&\bm J_{ij}^{(\ell)}(\bm r,t)
={}-G_{ij}\bm r_{ij}
\int_0^1 ds\,
W_\ell\!\left(
\bm r-\bm r_i-s\bm r_{ij}
\right)
\\
&\times
\int_0^1 ds'\,
\bm r_{ij}\cdot
\int d\bm r'\,
W_\ell\!\left(
\bm r'-\bm r_i-s'\bm r_{ij}
\right)\nabla' T_\ell(\bm r',t).
\end{aligned}
\label{eq:bondcurrent}
\end{equation}
The two smoothing operations play distinct roles: the first distributes the current carried by bond $(i,j)$ in space, while the second samples the coarse-grained temperature gradient driving that channel. In the scale-separated regime of Eq.~(\ref{eq:scales}), microscopic variations below $\ell$ are eliminated while hydrodynamic variations are retained, linking the discrete bond currents to the continuum constitutive relation below. Summing over all unordered pairs gives
\begin{equation}
\bm J_\ell(\bm r,t)
=
\sum_{i<j}
\bm J_{ij}^{(\ell)}(\bm r,t).
\label{eq:current_sum}
\end{equation}
Then, the resulting current can be written in the nonlocal constitutive form
\begin{equation}
\bm J_\ell(\bm r,t)
=
-\int_\Omega d\bm r'\,
\mathbb{K}_\ell(\bm r,\bm r')
\nabla' T_\ell(\bm r',t),
\label{eq:spatialnonlocal}
\end{equation}
where the coarse-grained transport kernel is
\begin{equation}
\begin{split}
\mathbb{K}_\ell&(\bm r,\bm r')
={}
\sum_{i<j}
G_{ij}\,
\bm r_{ij}\otimes\bm r_{ij}
\int_0^1ds\int_0^1ds'\\
&\times
W_\ell(\bm r-\bm r_i-s\bm r_{ij})
W_\ell(\bm r'-\bm r_i-s'\bm r_{ij}).
\end{split}
\label{eq:kernelmicro}
\end{equation}
Equation~(\ref{eq:kernelmicro}) provides an explicit microscopic-to-continuum map: $G_{ij}$ sets the weight of each radiative channel, while $\bm r_{ij}\otimes\bm r_{ij}$ encodes its range and orientation. The distinction between the interpolating microscopic field $T$ in Eq.~(\ref{eq:bondgradient}) and the coarse-grained field $T_\ell$ is essential; replacing microscopic gradients by $\nabla T_\ell$ is justified only after averaging over scales satisfying Eq.~(\ref{eq:scales}). The resulting constitutive relation is intrinsically nonlocal, since the current at $\bm r$ samples temperature gradients at spatially separated points $\bm r'$.

Energy conservation, by contrast, retains its local conservative form. If $u_\ell$ denotes the coarse-grained internal energy density,
\begin{equation}
\partial_tu_\ell(\bm r,t)
+
\nabla\cdot\bm J_\ell(\bm r,t)
=0.
\label{eq:energyconservation}
\end{equation}
Near $T_0$, we write
\begin{equation}
\delta u_\ell
=
c_V^{\rm eff}\delta T_\ell,
\end{equation}
where $c_V^{\rm eff}$ is the effective volumetric heat capacity. Combining Eqs.~\eqref{eq:spatialnonlocal} and \eqref{eq:energyconservation} gives
\begin{equation}
c_V^{\rm eff}\partial_tT_\ell(\bm r,t)
-
\nabla\cdot
\left[
\int_\Omega d\bm r'\,
\mathbb{K}_\ell(\bm r,\bm r')
\nabla'T_\ell(\bm r',t)
\right]
=0.
\label{eq:nonlocalhydro}
\end{equation}
This equation constitutes the central result of the present work: a nonlocal hydrodynamic description derived directly from the underlying microscopic radiative network. Its form is covariant under smooth coordinate transformations, with the nonlocal transport kernel transforming as a second-rank tensor density~\cite{Pendry2006}. This covariance opens a natural route toward extending transformation methods to nonlocal radiative transport, enabling the design of continuum media for steering, concentrating and redirecting radiative heat currents.

We next determine when this equation reduces to a local diffusion equation. In a statistically homogeneous medium sufficiently far from boundaries, the kernel becomes translationally invariant,
\begin{equation}
\mathbb{K}(\bm r,\bm r')
=
\mathbb{K}(\bm\xi),
\qquad
\bm\xi=\bm r'-\bm r,
\end{equation}
and Eq.~\eqref{eq:spatialnonlocal} becomes
\begin{equation}
\bm J(\bm r,t)
=
-\int d\bm\xi\,
\mathbb{K}(\bm\xi)
\nabla T(\bm r+\bm\xi,t).
\label{eq:convolution}
\end{equation}
If the kernel is sufficiently localized that its relevant spatial moments are finite, and if $L_T$ is large compared with its effective range, one may expand
\begin{equation}
\nabla T(\bm r+\bm\xi)
=
\nabla T(\bm r)
+
(\bm\xi\cdot\nabla)\nabla T(\bm r)
+\cdots.
\end{equation}
For a reciprocal statistically inversion-symmetric medium, the first spatial moment vanishes. The leading contribution is therefore
\begin{equation}
\bm J
=
-\bm\kappa_{\rm rad}\nabla T
+
O(\nabla^2T),
\end{equation}
with
\begin{equation}
\bm\kappa_{\rm rad}
=
\int d\bm\xi\,\mathbb{K}(\bm\xi).
\label{eq:Fourier}
\end{equation}
Thus Fourier's law emerges as the controlled long-wavelength limit of the nonlocal constitutive relation. Equivalently, on length scales much larger than the interaction range, the two-point kernel becomes effectively local
\begin{equation}
\mathbb{K}(\bm r,\bm r')
\longrightarrow
\bm\kappa_{\rm rad}\,
\delta(\bm r-\bm r').
\end{equation}
The existence of a local hydrodynamic limit requires not only the scale separation of Eq.~\eqref{eq:scales}, but also the convergence of the spatial moments entering the gradient expansion. To make this criterion explicit, we define the second spatial moment of the microscopic transfer network within a REV as
\begin{equation}
\mathbb{M}_2(\ell)
=\left\langle
\sum_j
\Gamma_{ij}
\bm r_{ij}\otimes\bm r_{ij}
\right\rangle_\ell .
\label{eq:M2}
\end{equation}
Its connection with diffusion follows by expanding the temperature field around $\bm r_i$,
\begin{equation}
\begin{aligned}
T(\bm r_j)
={}&T(\bm r_i)
+r_{ij,\alpha}\partial_\alpha T
&+\frac12 r_{ij,\alpha}r_{ij,\beta}
\partial_\alpha\partial_\beta T+\cdots .
\end{aligned}
\end{equation}
Substitution into Eq.~(\ref{eq:master}) shows that the first spatial moment generates a drift term, while the second generates diffusion. For an inversion-symmetric medium, opposite transfer directions are statistically equivalent, so that the first spatial moment vanishes. The leading contribution to the gradient expansion is therefore controlled by the second spatial moment,
\begin{equation}
\partial_t T
=\frac12\mathbb{M}_2:\nabla\nabla T+\cdots .
\end{equation}
For a homogeneous medium, this expression is equivalent to the anisotropic diffusion equation $\partial_tT=\nabla\cdot(\mathbb{D}_{\rm rad}\nabla T)$, with the radiative diffusion tensor directly determined by the second moment of the microscopic transfer network. The corresponding diffusivity and thermal conductivity are thus
\begin{equation}
\mathbb{D}_{\rm rad}
=\frac12\mathbb{M}_2,
\qquad
\bm\kappa_{\rm rad}
=c_V^{\rm eff}\mathbb{D}_{\rm rad}.
\label{eq:D}
\end{equation}

In an isotropic $d$-dimensional medium, all spatial directions are equivalent and the tensorial second moment is proportional to the identity, $\mathbb{M}_2=(\operatorname{Tr}\mathbb{M}_2/d)\mathbb{I}$. Since $\operatorname{Tr}(\bm r_{ij}\otimes\bm r_{ij})=r_{ij}^2$, the scalar radiative diffusivity becomes
\begin{equation}
D_{\rm rad}
=\frac{1}{2d}
\left\langle
\sum_j
\Gamma_{ij}r_{ij}^{2}
\right\rangle_{\rm REV}.
\label{eq:Diso}
\end{equation}
A finite local diffusivity therefore exists only if $\mathbb{M}_2(\ell)$ approaches an $\ell$-independent plateau within the scale-separated regime $\ell\ll L_T$. Otherwise, increasingly distant radiative channels continue to renormalize the effective transport coefficient, preventing the nonlocal kernel from being reduced to a local conductivity.

Figure~\ref{fig:M2} illustrates this criterion for one-dimensional metallic and polar nanoparticle chains. Fig.~\ref{fig:M2}(a) shows the pairwise radiative conductance $G(r)$ as a function of the separation $r$ from the central particle. For both materials, the long-distance behavior approaches $G(r)\sim r^{-2}$, revealing the algebraically long-ranged character of the underlying radiative network. Although the conductance associated with an individual distant channel decreases with separation, its contribution to the
second spatial moment is weighted by $r^2$. Consequently, for $G(r)\sim r^{-2}$, and hence $\Gamma(r)\sim r^{-2}$, distant channels remain relevant to the coarse-grained dynamics instead of becoming asymptotically negligible. This behavior is directly reflected in Fig.~\ref{fig:M2}(b), where $\mathbb{M}_2(\ell)$ grows approximately linearly with the coarse-graining length rather than approaching a plateau. In one dimension, the $r^{-2}$ decay of the transfer rate indeed yields $\mathbb{M}_2(\ell)\sim\ell$, in agreement with the numerical results. The absence of saturation therefore provides a direct signature that no scale-independent radiative diffusivity can be defined over the considered length scales. The two panels provide complementary microscopic and coarse-grained views of the same phenomenon: $G_{i,i+n}$ probes the spatial range of individual radiative channels, whereas $\mathbb{M}_2(\ell)$ quantifies their cumulative contribution to transport upon coarse graining. Fig.~\ref{fig:M2} thus shows explicitly how algebraically long-ranged microscopic interactions survive coarse graining and manifest themselves as persistent nonlocality at the hydrodynamic level.

More generally, the existence of a local hydrodynamic limit is controlled by the spatial decay of the microscopic radiative couplings, which can itself be strongly modified by the electromagnetic environment. Resonant cavities, for instance, can mediate long-distance energy transfer between spatially separated nanoemitters~\cite{Asheichyk2023}. To establish the corresponding locality criterion, consider a homogeneous $d$-dimensional network whose transfer rate decays asymptotically as $\Gamma(r)\sim r^{-\alpha}$. The number of channels contained in a radial shell scales as $r^{d-1}dr$, while each channel carries an additional factor $r^2$ in the second spatial moment. Introducing a large-distance cutoff $L$ therefore gives
\begin{equation}
\mathbb{M}_2(L)
\sim
\int^{L} dr\,r^{d+1-\alpha}.
\label{eq:powerlawmoment}
\end{equation}
The existence of a finite, scale-independent second moment in the limit $L\rightarrow\infty$ requires
\begin{equation}
\alpha>d+2.
\end{equation}
For $\alpha>d+2$, $\mathbb{M}_2$ converges and a local diffusive description can emerge at sufficiently large scales. At the marginal case $\alpha=d+2$, the second moment grows logarithmically, whereas for $\alpha<d+2$ it diverges algebraically as $L^{d+2-\alpha}$. In the latter two regimes, increasingly distant radiative channels continue to contribute to the transport coefficients, so that spatial nonlocality survives coarse graining and the continuum constitutive relation cannot, in general, be reduced to Fourier's law.

\begin{figure}[t]
\centering

\begin{tabular}{cc}
\includegraphics[width=0.5\columnwidth]{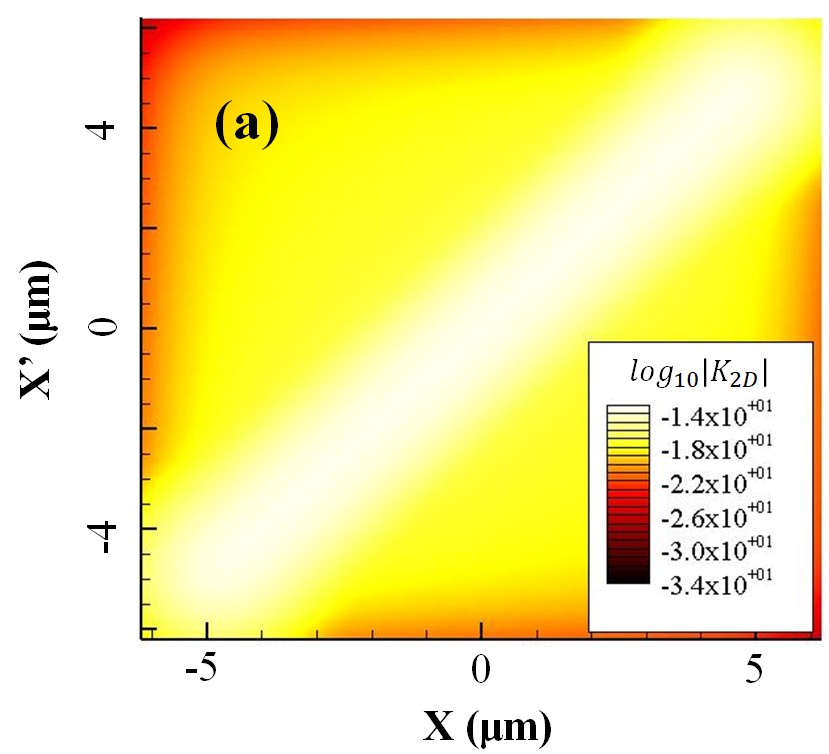} &
\includegraphics[width=0.48\columnwidth]{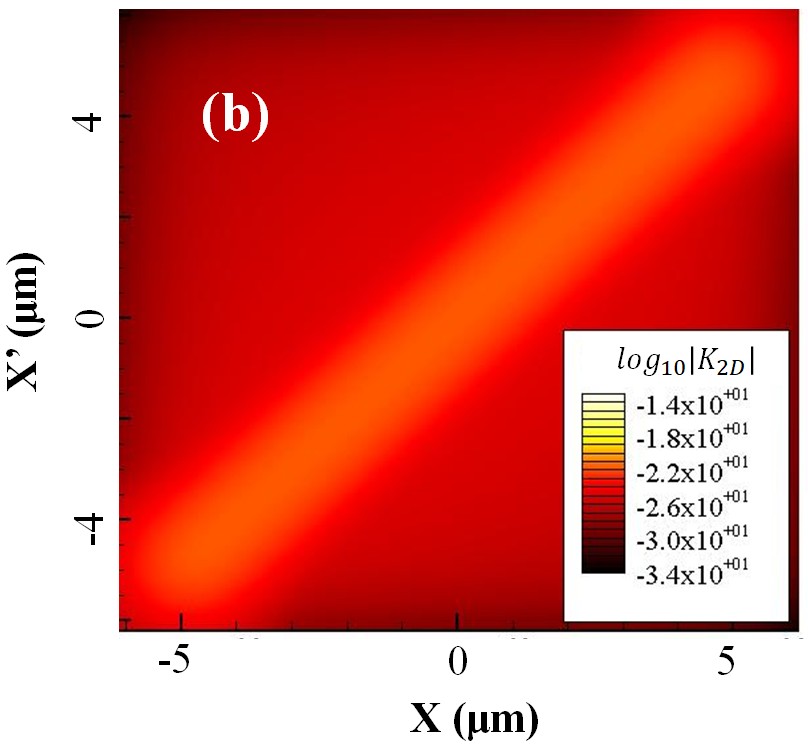}\\[3mm]
\multicolumn{2}{c}{\includegraphics[width=0.7\columnwidth]{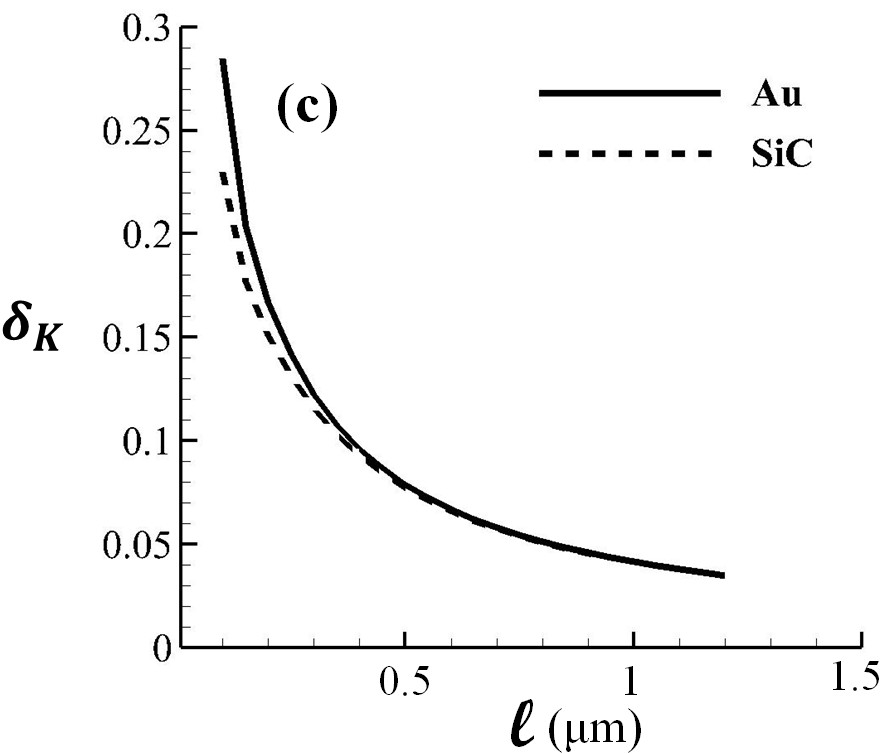}}
\end{tabular}

\caption{Spatial structure and convergence of the coarse-grained nonlocal transport kernel. (a,b) Maps of $K_\ell(x,x')$ for the SiC and Au chains, respectively; off-diagonal weight reveals long-range radiative channels. (c) Relative kernel variation $\delta_K(\ell)$ with
coarse-graining length. Its decrease demonstrates convergence toward an REV-independent continuum description without implying spatial locality.}
\label{fig:kernel}
\end{figure}

Beyond the moment criterion, the two-point kernel directly reveals the spatial structure of nonlocal transport. Figures~\ref{fig:kernel}(a) and \ref{fig:kernel}(b) show $K_\ell(x,x')$ for the SiC and Au chains, respectively. Its off-diagonal extension measures the separation between the position where the current is evaluated and where its driving temperature gradient is sampled. In SiC, these off-diagonal contributions are several orders of magnitude stronger than in Au, consistent with the surface-phonon-polariton-enhanced conductances in Fig.~\ref{fig:M2}(a). Thus, nonlocality is encoded both in the spatial extent of the kernel and in the transport weight carried away from its diagonal.

Convergence with the coarse-graining scale is quantified by
\begin{equation}
\delta_K(\ell)=
\frac{\|K_{\ell+\delta\ell}-K_\ell\|}{\|K_\ell\|},
\label{eq:deltaK}
\end{equation}
As shown in Fig.~\ref{fig:kernel}(c), $\delta_K$ decreases with $\ell$, demonstrating convergence toward an REV-independent continuum kernel. This does not imply locality: the limiting kernel may remain spatially extended and retain substantial off-diagonal weight.

The stationary consequences follow directly from the nonlocal constitutive law. For a chain of length $L$ held at $T(0)=T_L$ and $T(L)=T_R$, conservation requires $J(x)=J_{\rm ss}$, while
\begin{equation}
J_{\rm ss}=-\int_0^L dx'\,K(x,x')\,\partial_{x'}T(x').
\label{eq:chainss}
\end{equation}
Hence a uniform current need not imply a uniform temperature gradient. Writing $g(x)=\partial_xT(x)$ gives the Fredholm equation
\begin{equation}
\int_0^L dx'\,K(x,x')g(x')=-J_{\rm ss},
\label{eq:gradienteq}
\end{equation}
supplemented by $\int_0^L g(x)dx=T_R-T_L$, which determines $J_{\rm ss}$ and $g(x)$, and hence $T(x)=T_L+\int_0^x g(x')dx'$. In the local limit $K(x,x')\rightarrow\kappa_{\rm rad}\delta(x-x')$, one recovers $J_{\rm ss}=\kappa_{\rm rad}(T_L-T_R)/L$ and the linear Fourier profile. For an extended kernel, spatially separated gradients remain coupled, allowing non-Fourier stationary profiles even at constant heat current.

\begin{figure}[t]
\centering
\includegraphics[width=0.4\textwidth]{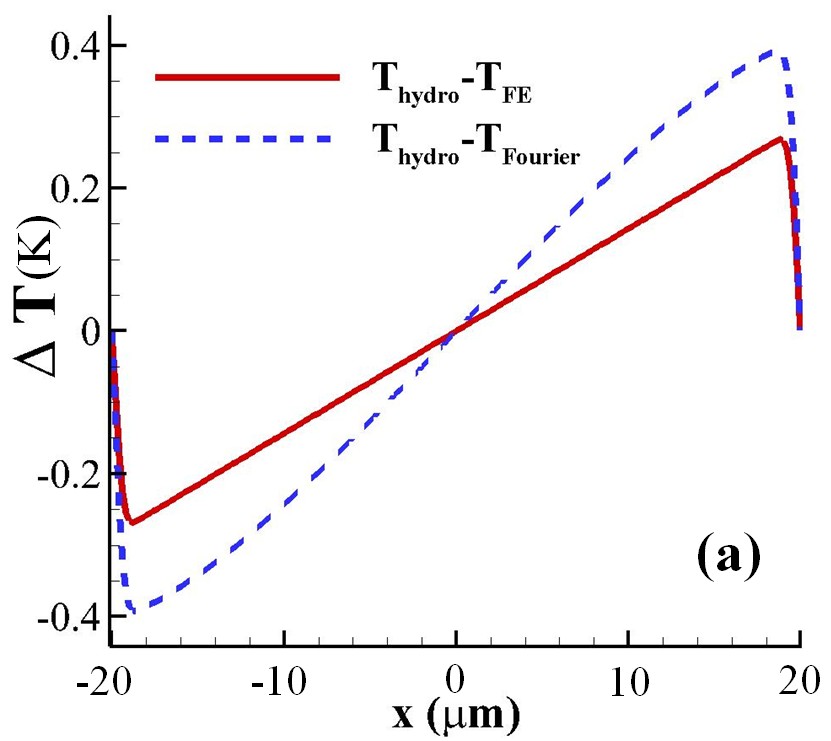}
\includegraphics[width=0.4\textwidth]{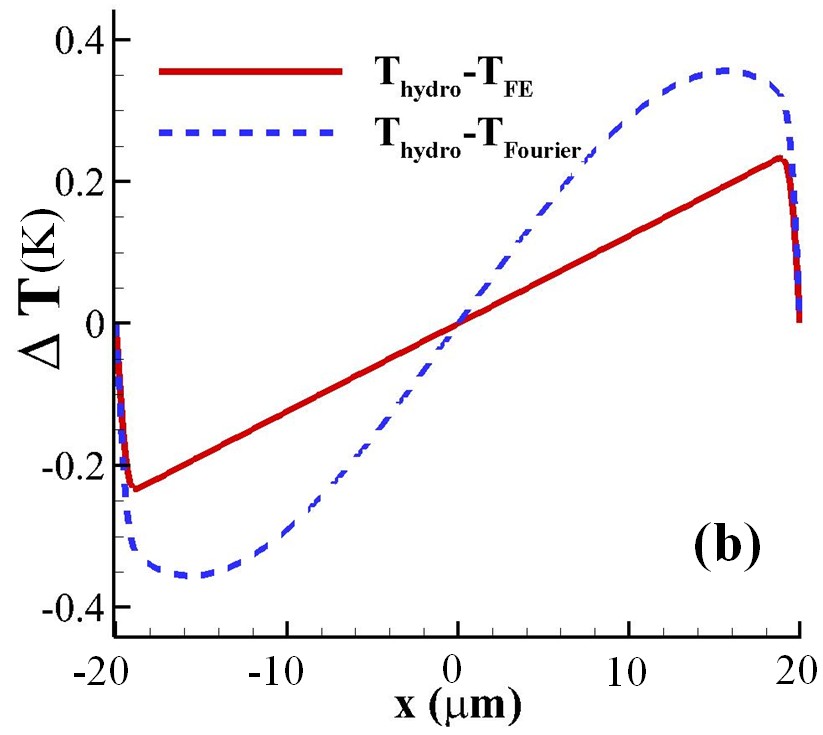}
\caption{Deviations from the stationary Fourier temperature profile for the SiC (a) and Au (b) nanoparticle chains. Results from discrete fluctuational electrodynamics (FE) are compared with the nonlocal hydrodynamic theory for $T_L=350$ K and $T_R=250$ K. The deviations vanish at the boundaries, consistently with the imposed temperatures, while their finite values inside the chains reveal the non-Fourier corrections induced by long-range radiative coupling.}
\label{fig:Tprofiles}
\end{figure}

Figure~\ref{fig:Tprofiles} directly compares the nonlocal continuum theory with the discrete fluctuational-electrodynamic dynamics. Since the stationary profiles remain nearly linear, we plot their deviations from the Fourier solution. Both microscopic FE and hydrodynamic results satisfy the imposed boundary temperatures, so the deviations vanish at the chain ends. In the interior, the hydrodynamic theory captures the magnitude and spatial dependence of the non-Fourier corrections, which are stronger in SiC owing to its enhanced long-range radiative coupling. Thus, an almost linear temperature profile does not imply local transport: nonlocality is revealed by the finite corrections generated by the off-diagonal structure of $K(x,x')$.

In summary, we have derived a nonlocal hydrodynamic description of many-body radiative heat transport directly from the underlying microscopic transfer network. Upon coarse graining, each radiative channel contributes to a two-point transport kernel through its conductance, spatial range and orientation, while local energy conservation is preserved. Conventional Fourier transport is recovered only when the relevant spatial moments of the microscopic transfer rates converge; otherwise, long-range radiative interactions persist at the continuum level as genuine transport nonlocality. This framework thus establishes a systematic bridge between microscopic fluctuational electrodynamics and continuum heat transport, enabling the treatment of large and geometrically complex many-body systems. More broadly, it provides a natural route to combine local phononic or electronic conduction within extended solids with nonlocal radiative exchange opening the way toward a unified continuum description of conductive and radiative heat transport.

\begin{acknowledgments}
P. B.-A. and R. M. acknowledge discussions with Matthias Krüger. F. H. acknowledges financial support by Deutsche Forschungsgemeinschaft under project number 570757245.
\end{acknowledgments}

\end{document}